# Current dependence of the hot-spot response spectrum of superconducting single-photon detectors with different layouts

I.Charaev[1], A. Semenov[2], S. Doerner[1], G. Gomard[3], K. Ilin[1], M. Siegel[1]

[1]*Institute of Micro- und Nanoelectronic Systems, Karlsruhe Institute of Technology (KIT), Hertzstrasse 16, 76187 Karlsruhe, Germany*

[2]*Institute of Optical Systems, German Aerospace Center (DLR), Rutherfordstrasse 2, 12489 Berlin, Germany*

[3]*Institute of Microstructure Technology (IMT), Karlsruhe Institute of Technology (KIT), Hermann-von-Helmholtz-Platz 1, 76344 Eggenstein-Leopoldshafen, Germany.*

E-mail: ilya.charaev@kit.edu

We show that avoiding bends in a current-carrying superconducting nanowire enhances the probability for low energy photons to be detected and that this enhancement is entirely due to the increase in the experimentally achievable critical current. We studied nanowires shaped as either meander or spiral. The spirals had different layouts, a double-spiral layout with an S-turn in the middle and a single-spiral layout without such turn. Nanowires were prepared from films of niobium nitride with a thickness of 5 nm. For specimens with each layout we measured the spectra of the single-photon response in the wavelength range from 400 nm to 1600 nm and defined the cut-off wavelength ($\lambda_c$) beyond which the response rolls off. The largest and the smallest $\lambda_c$ were found for the single-spiral layout and for the meander, respectively. For all three layouts the relationship between $\lambda_c$ and the relative bias current falls onto a universal curve which has been predicted earlier in the framework of the modified hot-spot model. For the single-spiral layout, the efficiency of photon detection at wavelengths smaller than $\lambda_c$ reaches the expected absorbance of the spiral structure and the timing jitter per unit length of the nanowire has the smallest value.

## 1. Introduction

Superconducting nanowire single-photon detectors have been the topic of intensive research over the last decade [1]. Many efforts have been taken to improve the detector design, particularly the layout, in order to increase the detection efficiency (DE) [2]. An enhancement of the detection efficiency is especially important at large wavelengths beyond the spectral cut-off where DE quickly degrades with the decrease of the photon energy. All hot-spot models of the detection mechanism [3], [4] predict larger cut-off wavelength for superconductors with smaller energy gap $\Delta$ ($\lambda_c^{-1} \propto \Delta^2$). Indeed, recent experiments with TaN [5] and WSi [6] meanders have demonstrated larger $\lambda_c$ than it was found for meanders from widely used NbN films.

Although it has been commonly agreed that $\lambda_c$ increases with the bias current $I_B$, the exact relationship between $I_B$ and $\lambda_c$ remains controversial [7]. According to the hard-core normal-spot model [8], [9] $1-I_B/I_C^d \propto \lambda_c^{-1/2}$, while the hot-spot model [3] predicts a linear relationship $1-I_B/I_C^d \propto \lambda_c^{-1}$, where $I_C^d$ is the de-pairing critical current. The numerical model invoking Ginsburg-Landau approach to the evolution of the order parameter [10] results in a $1-I_B/I_C^d \propto f(\lambda_c)$ dependence which falls between the two analytical relations above.

Since the relative current $I_B/I_C^d$ cannot be larger that $I_C/I_C^d$, where $I_C$ is the experimentally achievable critical current, the way to push $\lambda_c$ further towards larger values is to increase $I_C$. Despite of the advanced technology, in practical devices $I_C$ remains smaller than $I_C^d$ [7]. In NbN meanders, the ratio $I_C/I_C^d$ can be increased to some extent by variation of the film stoichiometry [11].

The major fundamental limitation on $I_C$ in meanders is imposed by the current-crowding effect [12] in the turns of the meander. The increased local current density in turns reduces locally the barrier height for vortex penetration and defines the critical current of the whole nanowire. According to the theory [12], the reduction factor of the critical current due to a bend in the nanowire depends on the angle of the bend, the wire width $w$, and the radius of the bend $r_b$. The effect is present in thin nanowires when $w \ll \Lambda = 2\lambda^2/d$, where $\Lambda$ is Pearl length and $\lambda$ is the magnetic penetration depth.

It has been suggested in Ref. [13] and has also been confirmed experimentally [14], [15] that external magnetic field with the proper orientation diminishes the effect of current crowding. This approach is not suitable for meanders due to alternating turns with opposite symmetry, but was successfully realized in specially designed square spirals [16] along with the enhancement of the photon count rate [17] in magnetic field. Additional benefit of operating photon detectors at larger bias currents is the decrease of the timing jitter [18]. The practical disadvantage of layouts with alternating straight wires and bends is their strong sensitivity to light polarization. The extinction factor increases with the increase in the wavelength [19].

Recently, a double-spiral layout was proposed as alternative design for single-photon detectors [20]. The absorption of spiral detectors is independent on polarization [21]. In comparison to meanders, specimens with the double-spiral layout demonstrated larger $I_C$, $\lambda_c$ and detection efficiency. Although the performance of photon detectors with the double-spiral geometry becomes better, the sharp S-bend in the current path through the center of the double-spiral layout still limits the experimental critical current of the specimen.

Here, we study the detection ability for single photons of an almost bend-free layout [16], [17] in the form of a single-spiral. We will compare three layouts, namely meander, double-spiral and single-spiral and show that the single-spiral layout offers the largest cut-off wavelength and smaller timing jitter. The preparation of samples will be described in the next section. Experimental results will be presented in sections 3 and 4. In section 5, we will analyze experimental results and compare them with theoretical models. We will conclude with a summary and a discussion of open issues in the last section.

2. **Preparation of nanowires**

Layouts of the specimens are shown in Fig. 1. A meander line fills the squared area 4 x 4 $\mu m^2$ (Fig. 1c). The spiral lines are both confined within the circle with a diameter of 7 $\mu m$. The area covered by spirals has a blind spot in the geometric center of the circle with a diameter either 1.5 $\mu m$ for the double spiral (Fig. 1 b) or 1.8 $\mu m$ for the single spiral (Fig. 1 a). For the single spiral the current flows between the contact outside of the circle and the contact at the blind spot. In the single-spiral layout, the largest bending radius of the current path equals the smallest radius of the spiral line. The double spiral layout includes two nested spirals which are connected at the blind spot. In this case, the current flows from the outer contact to the center of the circle and then back to another outer contact. The smallest bending radius of the current path in the double-spiral layout appears in its center and is less than one half of the smallest bending radius for single-spiral spiral layout (see Fig.1 b).

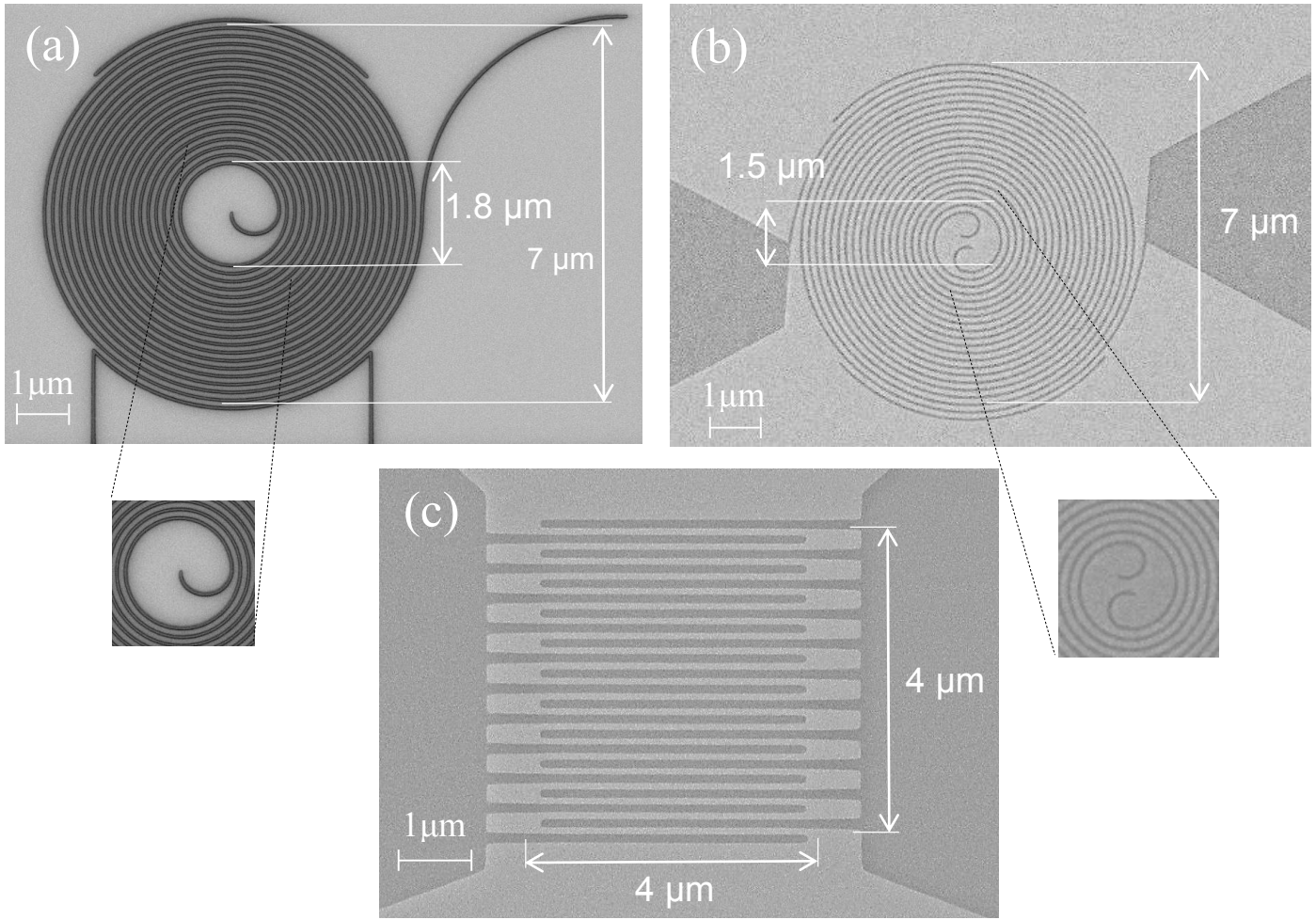

Fig.1. SEM images of specimens with different layouts: single-spiral (a), double-spiral (b), meander (c). The width of nanowires is the same for all layouts ($w$ = 100 nm). Dark areas correspond to the surface of the substrate (insulator); bright areas represent conducting NbN lines.

For all three layouts, the wired area terminates a coplanar line (not shown in Fig. 1) with an impedance of 50 Ω.

Fabrication process begins with the deposition of a thin NbN film on an R-plane cut, one-side polished, squared 10 x 10 mm² sapphire substrate. The films are grown by reactive magnetron sputtering of a pure Nb target in a gaseous mixture of argon and nitrogen with partial pressures $P_{Ar} = 1.8 \times 10^{-3}$ mbar and $P_{N2} = 3.5 \times 10^{-4}$ mbar of argon and nitrogen, respectively. During deposition the substrate was placed on a copper holder which in turn was placed on a heater. The latter was kept at a temperature of 850°C. The deposition rate of NbN was 0.08 nm/s at the discharge current of 160 mA. These conditions ensure the particular film stoichiometry which results in the highest critical temperature. We deposited films with a thickness of $d \approx 5$ nm, which was measured by a stylus profilometer.

Specimens with the meander and double-spiral layouts were prepared in two steps. First, the film was patterned by means of e-beam lithography and ion-beam etching to form contact pads and nanowires. The width $W$ of nanowires was 100 nm with a 100 nm gap in between, thus the geometric filling factor was 50%. The geometry was controlled by SEM inspection. Second, the coplanar line was made around the wired area by photolithography and reactive-ion etching. Similarly, a portion of the 100 nm wide straight line was patterned from the same film. The straight line was used for DC characterization of the films and, specifically, to obtain the reference value of the density of the critical current.

Preparation of specimens with the single-spiral layout is more complicated. Aside from pattering the film into spiral, it includes fabrication of an isolating layer between the spiral and the top contact and patterning the top contact itself. The nanowire in this layout had a width of 100 nm and was wound with a gap of 50 nm. Both determined the filling factor 66%. The 50 nm-thick AlN isolating layer was deposited by reactive-magnetron sputtering of Al target at room temperature. The 120 nm-thick Nb top electrode was sputtered in argon atmosphere without heating. Detailed information on the preparation of single-spirals can be found in [17].

From one film we fabricated ten specimens (3 pieces with each layout and one single bridge).

### 3. Characterization

#### A. Films

The temperature dependence of the film resistance was measured immediately after deposition in the range from 300 down to 4.2 K by standard four-probe technique. We defined the critical temperature $T_c$ as the lowest temperature at which a non-zero resistance could be measured. The measured critical temperature of the film was about 13 K. The resistivity, $\rho = R_S d$ was evaluated using the measured thickness, $d$, and the square resistance, $R_S$, of the films. The residual-resistivity ratio (*RRR*) is the ratio of the resistivity at room temperature to the maximum resistivity. The latter corresponds to the maximum square resistance of the film in the $R_S(T)$ dependence. For our film *RRR* = 0.96 and the resistivity at 20 K is 134 µΩ×cm. The temperature dependence of the second critical magnetic field $B_{C2}(T)$ was measured by applying an external magnetic field perpendicularly to the film surface. Using the theoretical dependence of $B_{C2}(T)$ in the dirty limit [22], we estimated the coherence length at zero temperature as

$$\xi(0) = \sqrt{\frac{\Phi_0}{2\pi B_{c2}}}, \tag{1}$$

that gave the coherence length of 4.6 nm for 5 nm thick film.
Calculation of the electron diffusion coefficient has been done by taking the linear part of the temperature dependence of $B_{C2}(T)$ near $T_c$ and the following formula

$$D = -\frac{4k_B}{\pi e}\left(\frac{dB_{c2}}{dT}\right)^{-1}. \tag{2}$$

This resulted in the value of electron diffusion coefficient $D = 0.55$ cm²/sec.

Using the experimentally measured superconducting energy gap [23] in NbN $\Delta(0) = 2.05\, k_B\, T_C$ we estimated the magnetic penetration depth as

$$\lambda(0) = \sqrt{\frac{\hbar\rho}{\pi\mu_0\Delta(0)}}. \tag{3}$$

We found λ(0) = 345 nm for our film.

Absorption spectra of a typical NbN film with a thickness 5 nm on the one-side polished sapphire substrate (thickness 0.4 mm) is shown in Fig. 2. The absorbance was obtained as 1- $t$ - $r$ where $t$ and $r$ are transmittance and reflectance, respectively. They both were measured with a spectrophotometer for free standing substrate with the film. The substrate with the film was illuminated either from the front side (the film) or from the rear side (substrate). For both configurations, the absorbance exhibits an increase in the range of wavelength from 400 nm to 850 nm and remains almost constant in the infrared range up to 1600 nm. The absorbance ratio for front side and rear side illumination remains almost constant at large wavelengths. It begins to decrease around 1000 nm that corresponds to the RMS roughness of the unpolished site of the substrate.

For measurements of the detection efficiency, specimens with the meander and the double spiral layouts were illuminated from the front side (right inset in Fig. 2). Because the nanowire in the single spiral layout was covered by an AlN isolating layer and a Nb-top electrode, which reflects light, the specimens with this layout were illuminated from the rear side, i.e., through the non-polished sapphire substrate. To eliminate the difference in the film absorbance for different illumination scenario and hence to avoid normalizing the measured detection efficiency for the absorbance, we used special flip-chip assembly shown in the left inset in Fig. 2. It is straightforward to show that for a thing film the film absorbance in two assemblies is the same with an accuracy better than 15%.

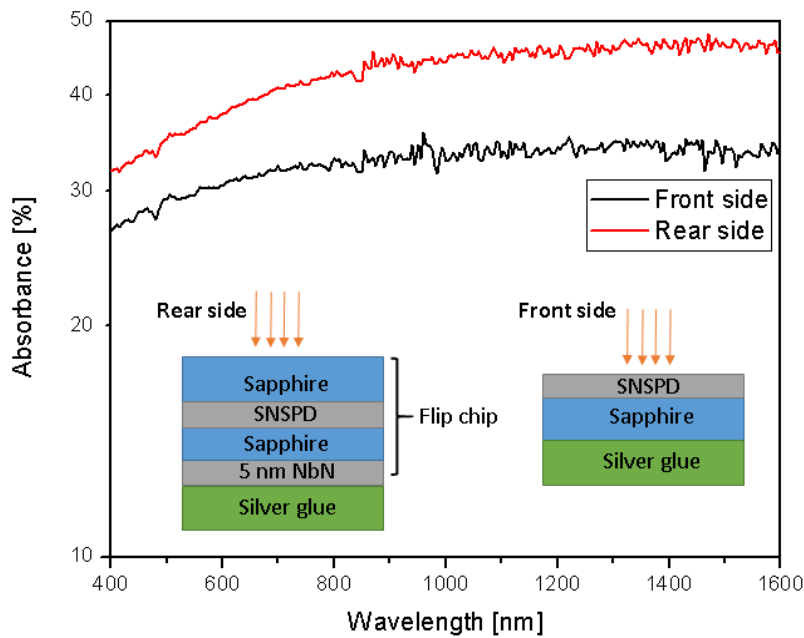

Fig.2. Absorption spectra of 5 nm thick NbN film on free standing one-side polished sapphire substrate illuminated from the rear side (upper curve) and from the front side (lover curve). Inset shows optical assemblies which were used for measurements of the detection efficiency.

### B. Specimens

After patterning the film into nanowire, the critical temperature of the nanowire becomes smaller than the $T_c$ of the film. Among specimens with different layouts the single bridge has the highest critical temperature 12.8 K. The slight variation of $T_c$ was found for meanders, double-spirals and single-spirals in the range between 12 and 12.4 K. The current-voltage (CV) characteristics of detectors were measured in the current-bias mode at 4.2 K. The critical current, $I_C$, of all structures was associated with the well-pronounced jump in the voltage from zero to a finite value corresponding to the normal state. The critical current density $j_c = \dfrac{I_c}{wd}$ of samples with different layouts is displayed in Fig. 3.

The highest value of 8.4 MA/cm² was obtained with the straight line. The critical current density of the meander was the lowest among these samples. The $j_c$ of meander is 38 % lower than the $j_c$ of the straight bridge with the same width. The decrease of the critical current density in spiral and meander nanowires is caused by a lower $T_c$ and by current crowding in bends.

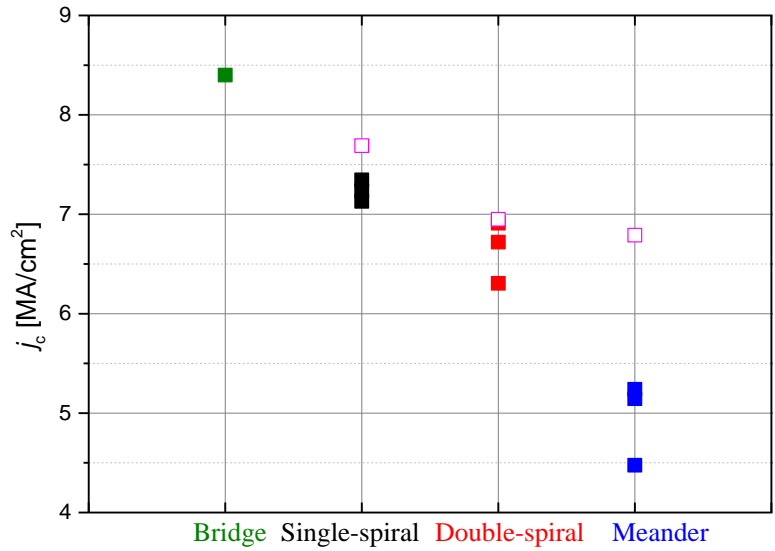

Fig.3. Critical-current densities $j_c$(4.2 K) of straight nanowires and nanowires with single and double spirals and meander-type layouts. The opened pink dots show the values scaled to the same relative temperature $T/T_C = 0.33$ for each layout.

To account for different critical temperatures, the critical current densities in Fig. 3 was scaled to the same relative temperature $T/T_C = 0.33$ by means of the Bardeen-temperature dependence of the critical current [24] (opened pink points in Fig. 3). Thus, the decrease of the critical-current density in meander due to its geometry is about half of the difference in the critical-current densities between the straight line and the meander at 4.2 K. The spirals demonstrate higher $j_c$ in comparison to the meander. Despite the absence of sharp bends in single spirals, their averaged $j_c = 7.2$ MA/cm² is below the value which was measured for a straight bridge. At the same time, $j_c$ of the double-spiral is slightly lower than the value of the single spiral.

To compare experimental critical currents with the theoretical limit, the de-pairing critical current $I_c^d$ was computed in the framework of the standard Ginzburg-Landau (GL) approach with the Bardeen-temperature dependence and the correction for the extreme dirty limit [25]:

$$I_c^d(T) = \frac{4\sqrt{\pi}\left(e^{\gamma}\right)^2}{21\varsigma(3)\sqrt{3}} \frac{\beta_0^2 (k_B T_c)^{\frac{3}{2}}}{eR_s\sqrt{D\hbar}} w\left[1-\left(\frac{T}{T_c}\right)^2\right]^{\frac{3}{2}} \times K\left(\frac{T}{T_c}\right), \tag{4}$$

where $\beta_0 = 2.05$ is the ratio of the energy gap at zero temperature to $k_B T_C$. The maximum ratio of the measured $I_c$ to the $I_c^d$ was 0.55, 0.48 and 0.4 for the single-spiral, double-spiral and meander layouts, respectively.

### 4. Experimental setup and results

Measurements were performed in a dipstick cryostat which was immersed in a standard ⁴He-transport dewar. The specimens were kept at an ambient temperature of 4.2 K. They were fixed on the sample holder together with temperature sensors and a bias tee and bonded with indium wires to contact pads on the holder. The low-temperature bias tee decoupled the high frequency path from the DC bias path. The high-frequency signal was led out of the dipstick by stainless-steel rigid coaxial cables, while DC bias was provided via a pair of twisted wires. The samples were biased by a battery-powered low-noise voltage source. The signal was amplified at room temperature by several amplifiers with the total gain of 70 dB and then sent to a pulse counter with a 300 MHz physical bandwidth.

The optical fiber feeds the light from the monochromator into the cryogenic part and is mounted on a movable stage above the sample surface. The distance between the end of the fiber and the surface of the detector is about 4 mm. This ensures that the size of the light spot on the specimen is much larger than the wired area. In this case, the distribution of the light intensity is homogenous across the wired area. The photon arrival rate to the illuminated side of the optical assembly (inset in Fig. 2) was carefully measured in the whole spectral range by means of a CCD camera and a sensitive photodiode.

#### A. Dark-Count Rate (DCR)

The dark-count rate DCR was determined by blocking the optical path at the warm edge. We found different behaviors of DCR for single spirals and other two types of SNSPD (Fig. 4).

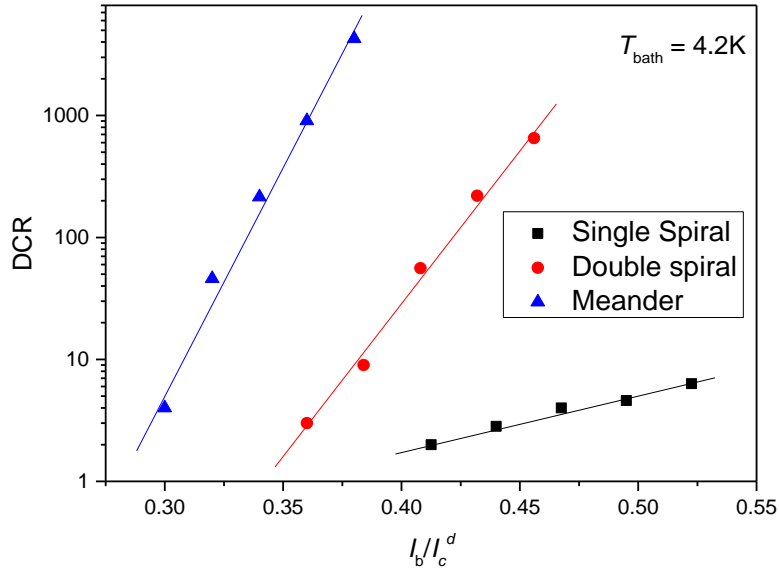

Fig.4. Dependence of the dark-count rate (DCR) on the relative bias current for three layouts which are indicated in the legend.

While DCR of the single-spiral SNSPD stayed below 10 cps for all relative currents, the double spiral and meander showed an increasing DCR up to $10^3$ cps at the relative bias current $0.95I_c$. We didn't observe dark counts at current lower than $0.75I_c$.

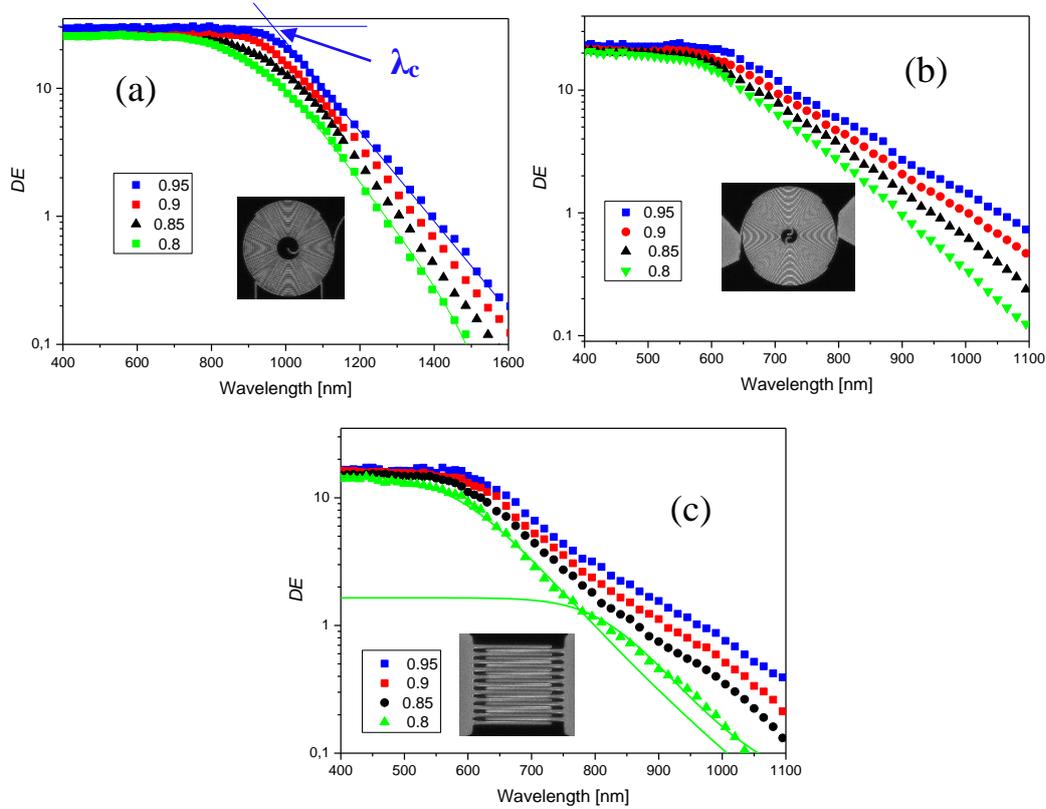

Fig. 5. Spectral detection efficiencies of specimens with single-spiral (a), double-spiral (b) and meander (c) layouts for different relative bias currents (shown in the legends). Solid green curves in panels (a) and (c) show best fits which were made with Eq. (6).

### A. Detection efficiency (DE)

The spectral detection efficiency for all types of detectors was measured at the same experimental conditions at different bias currents. For the particular wavelength, the DE was defined as the ratio of the rate of photon counts to the rate of photon arrival to the illuminated side of the specimen.

The spectral dependencies of the DE at different relative bias currents are shown in Fig. 5.

Each DE(λ) curve, demonstrates a clear roll off which begins around a particular wavelength. We call it the cut off wavelength $\lambda_C$ and define as the wavelength corresponding to the intersect of two straight lines which extrapolate the plateau at small wavelengths and the decaying portion of the DE(λ) curve (Fig. 5 a). At the same relative bias current ($0.95I_c$), the single spiral demonstrates $\lambda_C$ in the range of 900 nm, while the double spiral and the meander have smaller $\lambda_C$ (≈ 600 nm).

On the plateau, the detection efficiency of single spirals reaches 29.6%. Scaling this value to the filling factor of 50% of other layouts, we arrive at numbers close to the plateau efficiency 23.4 % of the double-spiral. In contrast, the meander has the plateau efficiency of only 15.7%.

When the bias current increases, DE grows much quicker at wavelengths beyond $\lambda_C$ than at the plateau. Correspondingly, $\lambda_C$ increases with the current.

## B. Timing jitter

To measure timing jitter, we used the experimental setup similar to the one described in [18]. Instead of a monochromator a 1560 nm femtosecond-pulsed laser (C-Fiber, Menlo Systems) was adopted as the single-photon source. The 32-GHz real-time oscilloscope builds a statistical distribution of the arrival times of photon counting pulses. The distribution typically has an almost Gaussian profile (Fig. 6). We defined the timing jitter as the full width at half maximum (FWHM) of the distribution. At the relative bias current $0.95 I_c$ we found the FWHM system jitter $j_{system}$ 49 ps, 38.6 ps and 42.7 ps for single-spiral, double-spiral and meander, respectively. The system jitter can be presented as the mid square of several components:

$$j_{system} = \sqrt{j_{SNSPD}^2 + j_{sync}^2 + j_{laser}^2 + j_{osci}^2} \quad , \tag{5}$$

where $j_{SNSPD}$, $j_{sync}$, $j_{laser}$, $j_{OSC}$ are the jitters from the specimen with the circuits, from the synchronization signal of the laser, from the laser itself and from the oscilloscope, respectively. Using (5) we obtained $j_{SNSPD}$ for single-spiral – 32 ps, double-spiral – 18.4 ps and meander – 22.5 ps. The jitter increases when the bias current decreases. This is illustrated in Fig. 7. Here, we normalized the jitter for the actual length of the nanowire in the particular design. The single-spiral has the smallest jitter per unit length and the weakest dependence of the jitter on the bias current. The jitter value and the steepness of this dependences increases for the double-spiral and further for the meander.

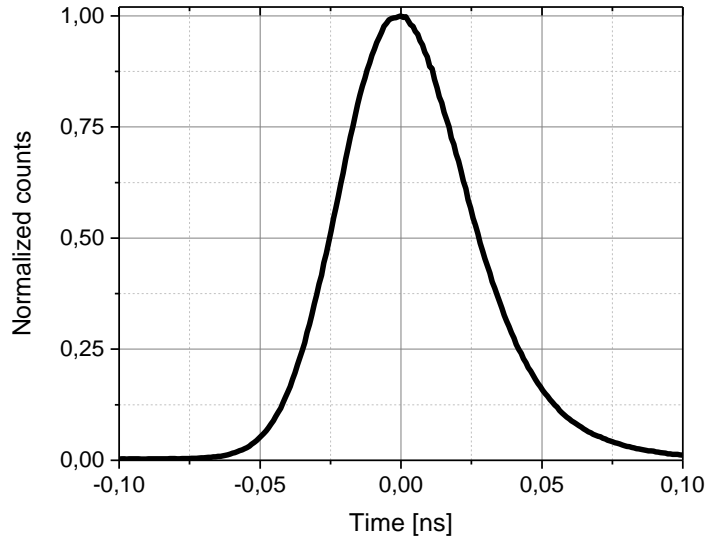

Fig.6. Distribution of the signal arrival times for the specimen with the single-spiral layout biased at $I_b = 0.95 I_c$. The jitter (FWHM of the distribution) amounts at 49 ps.

### 5. Discussion

We begin the analysis with the density of the critical current. It is commonly accepted [12], [26] that non-uniformities, constrictions and bends restrict the experimental critical current in nanowires. For the layouts studied here, there are turns in the meander and the sharp turn in the center of the double spiral where current crowding occurs. The straight nanowire which has no bends demonstrates the highest density of the critical current. Although the single-spiral layout is free from bends, $j_c$ remains smaller than in the straight nanowire. We speculate that the winding radius of the nanowire in the center of the spiral could be small enough to cause weak current-crowding effects.

The different rates of dark counts in our specimens support the envisaged role of turns [16]. Indeed, the meander contains turns with the smallest radius and has, correspondingly, the highest rate of dark counts. The smallest rate was observed in single spirals where the winding of the nanowire has the largest bending radius. The intermediate rate of dark counts in double spirals is easily attributed to the sharp turn of the current in the center of the layout where the nested spirals are connected to each other. We used the formalism of Bulaevskii et al. [27] to fit the dependences of DCR on bias current (Eqs. 51, 52 in Ref. [27]). The fits are shown with straight lines in Fig. 4. From the best fit we obtained the magnetic penetration depth at 4.2 K for all three studied layouts. The values 760 nm, 851 nm and 1767 nm for the meander, double and single spiral, respectively, are all noticeably larger that $\lambda = 346$ nm computed with Eq. 3. One of possible explanations can be that the approach of Ref. [27] does not apply to bends of any kind.

Our observation of the different timing jitter per unit length in specimens with different layouts supports the concept [28] that the nanowire can be thought of as a transmission line with its specific impedance $l/v$ where $l$ is the length of the nanowire and $v$ is the group velocity. Hence, the values shown in Fig. 7 are the reciprocal group velocity. Apart from the dispersion, the group velocity equals the phase velocity and depends on the inductance of the line per unit length $L$ as $v \propto L^{-0.5}$. For the superconducting transmission line, the inductance is primarily given by the kinetic inductance of the nanowire.

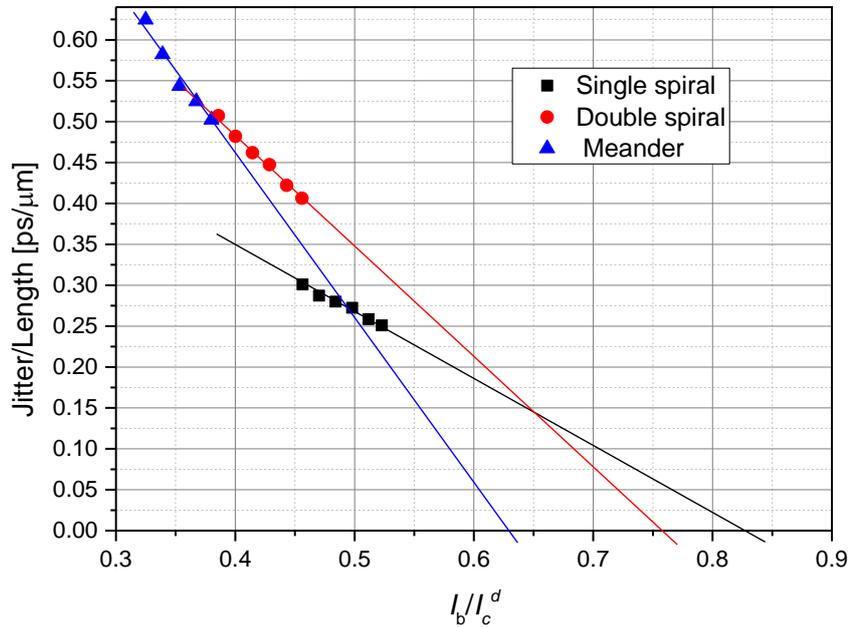

Fig.7. Timing jitters per unit length of different nanowires as function of the relative current for single-spiral, double-spiral and meander layouts.

We found earlier [29] that the total kinetic inductance of the meander grows with the current that qualitatively explains the current dependence of the timing jitter. Bends in the nanowire represent local disturbances in the transmission line which further reduce the average velocity and increase jitter. We speculate that the current crowding enhances the impact of the bias current on the local kinetic inductance in bends and that increases jitter in the meander as compared to single spirals.
The almost current-independent detection efficiency on plateaus for all three layouts (Fig. 5) represents the maximum efficiency when the whole nanowire together with turns contribute to the photon-count rate. The roll off begins when the central portion of the nanowire becomes inactive in detecting photons [30]. At even larger wavelengths, mostly bends and turns contribute to the photon-count rate [16]. The transition from the plateau to the decaying part of the DE spectrum is formally described by:

$$DE = \left(1 + \left(\frac{\lambda}{\lambda_c}\right)^p\right)^{-1}, \quad (6)$$

where $\lambda_c$ is the cut-off wavelength and the exponent $p$ describes the power-law decrease of the efficiency in the near infrared range. The solid curve in Fig. 5a shows the best fit of the spectral detection efficiency with the Eq. (6) which was obtained with $\lambda_C = 749$ nm.

The spectral detection effiecincy of the meander-type specimen shows a second knee for all relative bias currents (Fig. 5c). Beyond the cut-off wavelength, DE decreases monotonically untill at some point the DE jumps up (intersection of two solid curves in Fig. 5 c). We associate this knee with detection of photons by turns while the straight portions are less active than bends at wavelengths larger than at $\lambda = 760$ nm. Two solid curves depict fitting functions $DE_1(\lambda)$ and $DE_2(\lambda)$ which were obtained with equation (6). They represent the detection efficiency of straight and bended parts of meander, respectively. The detection effieciency of the meander-type specimen can be presented as:

$$DE(\lambda) = DE_1(\lambda) + DE_2(\lambda), \quad (7)$$

From the ratio of fitted detection effiecincies at palteaus of $DE_1(\lambda)$ and $DE_2(\lambda)$, we estimated the areal ratio of bends and straight parts of the meander. The equivalent number of squares in bended area amounts to 11% of the number of bends in straigt parts. This is very close to the geometric estimate of this ratio.

For all layouts the cut-off wavelength increases with bias current. In the framework of the model of the diffusiv hot-spot [3], the dependence is linear and reads

$$\lambda_c = \lambda_c(0)\left(1 - \frac{I_b}{I_c^d}\right)^{-1}. \quad (8)$$

$$\lambda_c(0)^{-1} = \frac{3\sqrt{\pi}}{4hc\varsigma}\frac{\Delta^2 w}{e^2 R_s}\sqrt{\frac{\tau}{D}}, \quad (9)$$

where $\tau$ – the electron thermalization time [31] and $\varsigma$ – is the effectiveness of the energy transfer from the absorbed photon to the electrons.

The steady-state hard-core model [8], [9] predicts a different dependence of the cut-off wavelength on the bias current:

$$\lambda_C = \frac{32 e^2 R_s D h c \varsigma}{\pi w^2 \Delta^2}\left(1 - \frac{I_b}{I_C^d}\right)^{-2} \quad (10)$$

For all three layouts we plot in Fig. 8 the relative bias current as function of the photon energy correspondig to the cut-off wavelength measured at this particular current along with best fits by three different models.

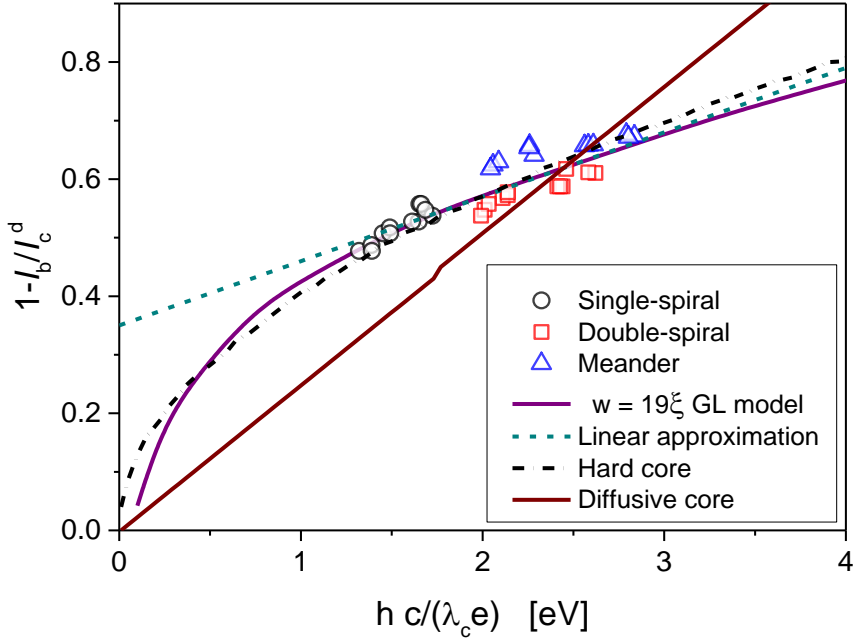

Fig.8 Inverted relative current $1-I_c/I_c^d$ versus energy of photons at the cut-off wavelength measured at this current. Symbols show experimental data for all three layouts. Lines show the best fits obtained with different theoretical models. The models are specified in the legend.

For all fits we used the effectivenes $\zeta$ as the only fitting parameter and the same superconducting and material parameters. The straight solid line represents the fit from the diffusion hot-spot model with $\zeta = 0.6$, the dash dotted line – from the hard-core model with $\zeta = 0.15$ and the curved solid line is the best fit from the numerical model invoking Ginsburg-Landau approach to the evolution of the order parameter [10] with $\zeta = 0.09$. For this fit we used the dimensionless wire width $w/\xi =19$ appropriate for our specimens. For comparison, the straight dashed line shows the phenomenological linear approximation from Ref. [32]. As it is clearly seen, the GL approach provides the best match with the experimental data simultaneously for all specimens. The fact that all experimental points fall on the same theoretical curve evidences the key role of the experimental critical current in the position of the spectral cut-off. This fact also confirms the supposition that the cut-off is caused by the position dependent probability of photons to be detected in a straight line and that it is almost unaffected by the bends in the layout.

## 6. Conclusion

For three different layouts of bended nanowires, we experimentally evaluated the performance parameters relevant for nanowire single-photon detectors. The major conclusion is that in terms of the detection efficiency and spectral cut-off the most important scaling factor is the experimental critical current for the particular layout while the bends play a minor role. Contrary, the rate of dark counts as well as the timing jitter are mostly restricted by the quality and number of bends in the layout. We have confirmed the supposition that for timing jitter the nanowire acts as a transmission line and that the amount of timing jitter is directly related to the length of the nanowire, its kinetic inductance and the bends.

ACKNOWLEDGMENT

I.C. acknowledges support from Karlsruhe School of Optics and Photonics of Karlsruhe Institute of Technology. G.G. acknowledges the support of the Helmholtz Postdoctoral Program.